\documentclass[aps,prl,twocolumn,showpacs]{revtex4}
\usepackage{amsmath,amssymb}

\begin{document}
\title{Quantum Byzantine Agreement with a Single Qutrit}
\author{Mohamed Bourennane}
\email{boure@physto.se}
\affiliation{Department of Physics, Stockholm University,
SE-10691 Stockholm, Sweden}
\author{Ad\'an Cabello}
\email{adan@us.es}
\affiliation{Departamento de F\'{\i}sica Aplicada II, Universidad de
Sevilla, E-41012 Sevilla, Spain}
\author{Marek \.{Z}ukowski}
\email{marek.zukowski@univie.ac.at}
\affiliation{Institute of Theoretical Physics and Astrophysics,
Uniwersytet Gda\'nski, PL-80-952 Gda\'nsk, Poland}


\date{\today}


\begin{abstract}
Quantum mechanics provides several methods to generate and
securely distribute private lists of numbers suitably
correlated to solve the Three Byzantine Generals Problem. So
far, these methods are based on three-qutrit singlet states,
four-qubit entangled states, and three or two pairwise quantum
key distribution channels. Here we show that the problem can be
solved using a single qutrit. This scheme presents some
advantages over previous schemes, and emphasizes the
specific role of qutrits in basic quantum information processing.
\end{abstract}



\maketitle


It has been recently shown that optimal quantum solutions for
some multiparty communication tasks do not require
entanglement. Protocols using only the sequential communication
of a single qubit have been demonstrated for secret sharing
\cite{STBKZW05} and some communication complexity problems
\cite{TSBBZW05}. These protocols were shown to be much more
resistant to noise and imperfections than previous protocols
based on entanglement. Here we shall present a new example of a
problem which can find an optimal quantum solution in the form
of a sequential exchange of a single quantum system.

The Three Byzantine Generals Problem (TBGP) expresses
abstractly the problem of achieving coordination between the
nonfaulty components of a distributed computation when some
components fail \cite{PSL80,LSP82}. Three divisions of the
Byzantine army, each commanded by its own general, are
besieging an enemy city. The three generals, Alexander,
Buonaparte, and Clausevitz ($A$, $B$, and $C$) can communicate
with one another by messenger only (i.e., by pairwise
authenticated error-free classical channels). They must decide
upon a common plan of action: either to attack ($0$) or to
retreat ($1$). The commanding general $A$ decides on a plan and
communicates this plan to the other two generals by sending $B$
a message $m_{AB}$ (either $0$ or $1$), and by sending $C$ a
message $m_{AC}$. Then, $B$ communicates the plan to $C$ by
sending him a message $m_{BC}$, and $C$ communicates the plan
to $B$ by sending him a message $m_{CB}$. However, one of the
generals (including $A$) might be a traitor, trying to keep the
loyal generals from agreeing on a plan. The TBGP is to find a
way in which: (i) all loyal generals follow the same plan, and
(ii) if $A$ is loyal, then every loyal general follows the plan
decided by $A$.

The TBGP is unsolvable \cite{PSL80,LSP82}, unless the generals share
some suitable private data. Each of the generals must be in
possession of a list of numbers unknown to the other generals, but
suitably correlated with the corresponding lists of the other
generals. There is no method, neither classical nor quantum, to
guarantee the success of the distribution of the required lists.
Nevertheless, a variation of the TBGP, called Detectable Byzantine
Agreement (DBA) or Detectable Broadcast \cite{FGM01,FGHHS02}, which
is unsolvable by classical means \cite{FLM86}, can be solved using
quantum resources
\cite{FGM01,FGHHS02,Cabello02,Cabello03a,Cabello03b,IG05,GBKCW07}.
In the DBA, conditions (i) and (ii) are relaxed so (i') either all
loyal generals follow the same plan or all abort, and (ii') if $A$
is loyal, then either every loyal general follows the plan decided
by $A$ or aborts.


\begin{table*}[htb]
\caption{\label{tab1}At the beginning of step (ii), $C$ has
received $m_{AC}$ in the form of $l_{AC}$ from $A$, and
$m_{BC}$ in the form of $l_{BC}$ from $B$. The table shows what
$C$ will do, depending on what $C$ obtains when he checks the
consistency between these data and his own list $l_C$.
$\{m_{AC},l_{AC}\} \cong l_C$ means that $m_{AC}$ and $l_{AC}$
are found to be consistent with $l_C$, $\not\cong$ means
``inconsistent with,'' and $\bot$ means ``I have received
inconsistent data.''}
\begin{ruledtabular}
{\begin{tabular}{c|l|l}
& {\rm If} & {\rm then $C$ will follow the plan} \\
\hline
(iia) & $\{m_{AC},l_{AC}\} \cong l_C\,\&\,\{m_{BC},l_{BC}\} \cong l_C\,\&\,m_{AC} = m_{BC}$ & {\rm $m_{AC}=m_{BC}$ (no traitor)} \\
(iib) & $\{m_{AC},l_{AC}\} \cong l_C\,\&\,\{m_{BC},l_{BC}\} \cong l_C\,\&\,m_{AC} \neq m_{BC}$ & {\rm previously decided by $B$ and $C$ ($A$ is the traitor)} \\
(iic) & $\{m_{AC},l_{AC}\} \cong l_C\,\&\,m_{BC}=\bot$ & {\rm $m_{AC}$ (although $A$ can be the traitor)} \\
(iid) & $\{m_{AC},l_{AC}\} \cong l_C\,\&\,\{m_{BC},l_{BC}\} \not\cong l_C$ & {\rm $m_{AC}$ ($B$ is the traitor)} \\
(iie) & $\{m_{AC},l_{AC}\} \not\cong l_C\,\&\,\{m_{BC},l_{BC}\} \cong l_C$ & {\rm $m_{BC}$ ($A$ is the traitor)} \\
(iif) & $\{m_{AC},l_{AC}\} \not\cong l_C\,\&\,m_{BC}=\bot$ & {\rm previously decided by $B$ and $C$ ($A$ is the traitor)} \\
\end{tabular}}
\end{ruledtabular}
\end{table*}


Quantum mechanics provides several methods to generate and securely
distribute the required lists. So far, these methods are based on
three-qutrit singlet states \cite{FGM01,Cabello02,Cabello03a},
four-qubit entangled states \cite{Cabello03b,GBKCW07}, and three
\cite{FGHHS02} or two \cite{IG05} pairwise quantum key distribution
(QKD) channels. In this Letter we introduce a protocol to generate
and securely distribute these lists using a single qutrit. We assume
that $A$, $B$, and $C$ can communicate with one another by pairwise
authenticated error-free classical channels and pairwise
authenticated quantum channels.


{\em Correlated lists and their use.---}The goal of the
protocol is to distribute three lists, $l_{A}$ known only by
$A$, $l_{B}$ known only by $B$, and $l_{C}$ known only by $C$,
all of the same length $L$, with the property that if $0$ ($1$)
is at position $j$ in $l_{A}$, then $0$ ($1$) is at position
$j$ both in $l_{B}$ and in $l_{C}$, and if $2$ is at position
$j$ in $l_{A}$, then $0$ is at position $j$ in one of the other
lists and $1$ is at position $j$ in the other. The combinations
$201$ and $210$ occur with the same probability
\cite{Cabello03b, GBKCW07}.

Before we proceed further, note that, on one hand, $A$ knows
exactly at which positions the lists $l_A$ and $l_B$ are
perfectly correlated, and at which positions they are
anticorrelated (but in this case he has no faintest idea who
has $1$ and who has $0$). On the other hand, $B$ and $C$ do not
know whether their data at a given position are correlated or
anticorrelated.

Once the parties have these lists, they can use them to reach an
agreement following a protocol introduced in \cite{GBKCW07} and
summarized here for completeness' sake:

(i) When $A$ wants to send $B$ a message $m_{AB}$ (attack, $1$,
or retreat, $0$) he sends $B$ a list $l_{AB}$ of all of the
positions in $l_{A}$ in which the value $m_{AB}$ appears. After
that, if $A$ is loyal he will follow his plan of military
action.

The roles of $B$ and $C$ are symmetrical, and thus everything
we say about $B$ applies to $C$ and vice versa. When $B$
receives $m_{AB}$ in the form of $l_{AB}$, only one of two
things are allowed to happen:

(ia) If $l_{AB}$ is of the appropriate length (i.e.,
approximately $L/3$), and $l_{AB}$, and $l_{B}$ are consistent
at each position $j$ (i.e. they fulfill the property of the
lists), then $B$ will follow the plan $m_{AB}$ implied by the
received $l_{AB}$ {\em unless} $C$ convinces him that $A$ is
the traitor in the next step [see (ii)].

(ib) If $l_{AB}$ and $l_{B}$ are inconsistent, then $B$
ascertains that $A$ is the traitor and $B$ will not follow any
plan until he reaches an agreement with $C$ in the next step
[see (ii)].

(ii) The message $m_{BC}$ $B$ sends $C$ can be not only $0$ or
$1$, but also $\bot$, meaning ``I have received inconsistent
data.'' If the message to be conveyed, $m_{AC}$ is $0$ or $1$,
$B$ sends $C$ a list $l_{BC}$ which is, if he is loyal and so
is $A$, the same list $l_{AB}$ that $B$ has received from $A$.
After $C$ receives $m_{BC}$ in the form of $l_{BC}$, he
compares it with $m_{AC}$ received earlier in the form of
$l_{AC}$. Then, only one of six situations can happen, which
are listed in the Table \ref{tab1}.


{\em Quantum distribution of the lists.---}Now we shall explain the
quantum protocol for distributing these lists. The three generals
have devices which can unitarily transform qutrits. In addition,
general $A$ has also a source of qutrits and general $C$ has a
detection station. The three generals act according to the following
protocol.

(I) {\em Initial state.} $A$ prepares his qutrit in the state
\begin{equation}
|\psi_0\rangle=\frac{1}{\sqrt{3}}(|0\rangle+|1\rangle + |2\rangle).
\label{state}
\end{equation}

(II) {\em $A$ sets the basis.} His first choice is the ``basis
choice'', which is to decide whether he will be coding his number in
the basis $I$, in which case he does not perform any initial unitary
transformation, or in the basis $II$, in which case he acts with a
unitary operator
\begin{equation}
U_{II}=\frac{1}{\sqrt{3}}\left(\begin{array}{ccc} 1 & 0 & 0 \\
0 & e^{i\,2 \pi /3} & 0 \\
0 & 0 & e^{i\,2 \pi /3} \end{array}\right).
\end{equation}
Note that, from the interferometric point of view, the type $II$
operation has no effects on the beam $0$ and introduces a phase
shift by $2 \pi/3$ in the beams $1$ and $2$.

(III) {\em $A$ encodes the number.} The next choice of $A$ is to
encode one of the three random numbers $0$, $1$, and $2$ (all with
the same probability). If he wants to encode $n$, he performs
\begin{equation}
U(n)=\frac{1}{\sqrt{3}}\left(\begin{array}{ccc} 1 & 0 & 0 \\
0 & e^{i\,2 n \pi /3} & 0 \\
0 & 0 & e^{-i\,2 n \pi /3} \end{array}\right).
\end{equation}
After that, the qutrit is sent to general $B$.

(IV) {\em $B$ chooses the basis.} $B$ either performs $U_{II}$ (type
$II$ basis encoding) or does nothing (type $I$).

(V) {\em $B$ encodes the number.} $B$ is allowed to encode $0$ or
$1$ with equal probabilities. If he is to encode $0$, he does
nothing, that is $U(0)$, in the other case he acts with $U(1)$. He
then sends the qutrit to general $C$.

(VI) {\em $C$ chooses the basis and encodes the number} in exactly
the same way as $B$.

(VII) {\em $C$ measures the qutrit} using a device that
distinguishes the state $|\psi_0\rangle$ given by \eqref{state}
from any two other states orthogonal to it, e.g., an unbiased
multiport beamsplitter.

(VIII) If $C$ gets $|\psi_0\rangle$, the generals reveal their {\em
bases}, but not the encoded numbers. They do this in reverse order:
first $C$, last $A$. If it turns out that all of them chose to do
nothing, or all of them chose to perform $U_{II}$, the run is
treated as a valid distribution of the secret numbers.

The protocol distributes the numbers in the required way because
\begin{equation}
U_{II}^3=\openone,\end{equation} where $\openone$ is the identity matrix,  and 
\begin{equation}
U(k)U(l)U(m)=\openone,\end{equation} {whenever $k+l+m=0$ modulo 3.}


{\em Security.---}The traitor general may be eavesdropping
either by an intercept-resend method or by entangling the
qutrit with another system. However, any of these attacks cause
a disturbance which can be detected in a random check of some
of the valid runs (exchanging the actual numbers between the
parties). To prove that the quantum distribution of the lists
is secure against these attacks, note that, if there is no
eavesdropping, the state that the last general receives must be
a pure state. After all the generals reveal their bases (in the
right time order, first $C$, next $B$, and finally $A$), and if
the bases are the same, then the final state must be an
eigenstate of the final measurement basis, namely
$|\psi_0\rangle$. Therefore, the measurement results should be
perfectly deterministic because the state of the qutrit is pure
at any step of the protocol. However, any eavesdropping by the
traitor general, which is about to give him information about
the state of the qutrit, as it is done before the bases are
revealed, must lead to correlations with classical (e.g., in
the intercept-resend strategy) or quantum (e.g., when
entangling the qutrit with an ancilla) states of some systems
monitored by the traitor. That is, the effective state
(averaged over the selected runs in which all the generals
claim that they set the same basis) reaching the final
measurement station is a mixed state. There does not exist any
observable for which a mixed state gives a deterministic
prediction. 

In the protocol   parts of the lists must be revealed for the
cross-check for eavesdropping or cheating. If one requires that the order of revealing the
numbers is random, then the traitor will not be the last one to announce his value in approximately $2/3$ of cases. In
such cases, the traitor has no way to announce a number which
is always consistent. Thus from the analysis of the errors, the
loyal generals can conclude that there was cheating. Note that,
paradoxically, in order not to be uncovered in those cases when
the traitor is chosen to be the last one to declare his number,
he must give, from time to time, an inconsistent number (i.e.,
not fulfilling the $(k+l+m)_{\rm{mod}\,3}=0$ rule). Otherwise,
the set of cases in which he reveals last would look suspiciously
perfect.

Furthermore, note that $C$ could also cheat when announcing in
which runs he received measurements consistent with
$|\psi_0\rangle$. But then, the cheating is easily detectable
after the bases are revealed, since {\em only} valid runs
deterministically lead to $|\psi_0\rangle$. This is because
$U_{II}^2\neq\openone$, etc.

After this protocol, each of the parties has a final list. If all the
results are correctly correlated, the generals would assume that the
remaining results are correctly correlated and will use the
resulting lists $l_{A}$, $l_{B}$ and $l_{C}$ to reach an agreement,
as we explained before. In case of failure of this part, the loyal
generals agree to abort.


{\em Possible experimental implementations.---}The single
qutrit required for the protocol can be realized in many ways.
One of them would be by means of the unbiased multiport
beamsplitters \cite{ZZH97}. Other possibility is time-bin
\cite{MRTSZG02} realization of qutrits. Furthermore, one can
use type-II spontaneous parametric down-conversion and treat
the three symmetric two-photon polarization states as the basis
state of a composite qutrit \cite{BCKMZOT04}. Finally, another
possibility is using single photons passing trough a triple
slit \cite{SCMSLSW09}.


{\em Advantages over QKD protocols.---}The single-qutrit scheme has
two main advantages versus the the scheme in \cite{IG05}:

(i) The scheme in \cite{IG05} consists of two QKD channels. Each of
them requires the preparation and the measurement of qubits.
Therefore, a successful distribution of one number of the lists
requires at least two detections. Indeed, it requires four
detections if the QKD is based on von Neumann measurements on a
single qubit, since each QKD channel must transmit a trit value. If
the efficiency of the detectors $\eta$ is not perfect, then a
successful distribution occurs with only probability $\eta^2$ (more
realistically, only with probability $\eta^4$). In the single-qutrit
scheme a successful distribution occurs with probability $\eta$. The single qutrit method
scales much more efficiently with a growing number of generals. This makes such a scheme 
even more favorable. 

(ii) The goal of the scheme in \cite{IG05} is to distribute
lists of six combinations of numbers (0--1--2, 0--2--1,
1--0--2, 1--2--0, 2--0--1, and 2--1--0). The goal of the
single-qutrit scheme is to distribute simpler lists with a
different symmetry; lists of four combinations of numbers
(0--0--0, 1--1--1, 2--0--1, and 2--1--0). The classical part of
single-qutrit scheme is therefore more efficient than that of
the scheme in \cite{IG05}.

{\em Conclusions.---}Single qutrits allow QKD protocols with
additional security features \cite{BP00, Svozil09b}, quantum
random number generation \cite{Svozil09a}, and
better-than-classical performance in games which require
entanglement when they are played with two qubits \cite{AV08}.
Here we have presented the first application of single qutrits,
which provides an optimal quantum solution to a multiparty
communication problem. All these results suggest that the
qutrit provides a very specific quantum resource which is positioned between the
simplest quantum superposition, represented by the qubit, and the simplest form of entanglement, represented by the
two-qubit entanglement.




\begin{acknowledgments}
The authors thank N. Gisin, C. Kurtsiefer, and H. Weinfurter
for useful conversations. This work was supported by the EU 6FP
programmes QAP and SCALA, and the Swedish Research Council
(VR). A.C. acknowledges support from the Spanish MEC Project
No. FIS2008-05596, and the Junta de Andaluc\'{\i}a Excellence
Project No. P06-FQM-02243. M.\.{Z}. was supported by Wenner Gren
Foundation.
\end{acknowledgments}


\end{document}